\documentclass[preprint,showpacs,showkeys,preprintnumbers,amsmath,amssymb,aps,pra]{revtex4-1}

\usepackage{bm,amssymb,amsmath,mathrsfs}
\usepackage[usenames]{color}
\usepackage{enumerate}
\usepackage{bookmark}

\usepackage{dcolumn}
\usepackage[pdftex]{graphicx}

\newcommand{\bc}{\begin{center}}
\newcommand{\ec}{\end{center}}
\newcommand{\bit}{\begin{itemize}}
\newcommand{\eit}{\end{itemize}}
\newcommand{\bq}{\begin{equation}}
\newcommand{\eq}{\end{equation}}

\begin{document}

\preprint{CC 15\_9}

\title{Decay Lifetimes of the Vector and Tensor Polarization of a Stored Deuteron Beam}

\author{S.~R.~Mane}
\email{srmane001@gmail.com}

\affiliation{Convergent Computing Inc., P.~O.~Box 561, Shoreham, NY 11786, USA}

\date{\today}

\begin{abstract}
An rf solenoid was operated in the IUCF Cooler
to induce a depolarizing spin resonance 
for a stored polarized deuteron beam.
The decay lifetimes of the vector and tensor polarizations
were found to be in the ratio $1.9:1$,
as opposed to the expected ratio of $3:1$ from standard angular momentum theory.
This report describes my investigations and attempts to
formulate a theoretical explanation of the above phenomenon.
\end{abstract}



\maketitle


\baselineskip 4ex


Approximately in 2005, my dear friend Waldo Mackay brought the following paper to my attention \cite{PolLifetime_PRE2003}.
The paper described experimental measurements of the decay rates of the vector and tensor polarizations
of a stored polarized deuteron beams circulated in the IUCF Cooler.
The beam was stored at flattop and a depolarizing spin resonance was induced by operating an rf solenoid
at a fixed frequency close to the spin precession frequency.
There was no Siberian Snake in the IUCF Cooler for these studies; 
the spins (on the reference orbit) precessed around the vertical axis (in the absence of the rf solenoid).
The IUCF Cooler contained a so-called ``partial Type 3 Siberian Snake,'' which shifted the spin tune slightly,
but that was accounted for in the studies reported in \cite{PolLifetime_PRE2003}.

It was reported in \cite{PolLifetime_PRE2003} that
the observed ratio of the vector to tensor polarization decay lifetimes was $1.9:1$,
whereas (as stated in the paper) na{\"\i}ve angular momentum theory says the ratio should be $3:1$.
Waldo had some ideas as to why this might be so, based on work I had published in my Ph.D.~thesis
and he invited my collaboration to analyze the problem and formulate a theoretical explanation of the data.
Regrettably, I did not take up his kind invitation.
Recently, the paper \cite{PolLifetime_PRE2003} came to mind again.

The basic details of the studies in \cite{PolLifetime_PRE2003} are as follows.
Following \cite{PolLifetime_PRE2003}, the vector and tensor polarizations will be denoted by $P_z$ and $P_{zz}$ below.
A polarized deuteron beam was injected into the IUCF Cooler and accelerated to a kinetic energy of 270 MeV and cooled.
An rf solenoid was then switched on, to deliberately induce a depolarizing spin resonance.
Note that the betatron tunes were adjusted to avoid the $G\gamma+5 = \nu_y$ resonance.
(The authors denoted the vertical betatron tune by $\nu_y$.)
To minimize systematic errors, the injected beam was cycled through multiple states of vector and tensor polarization,
listed in Table I in \cite{PolLifetime_PRE2003}.

The polarization measurements consisted of two overall parts.
In the first part, the precise location of the center of the spin resonance was determined.
The text below Fig.~3 in \cite{PolLifetime_PRE2003} states 
the rf solenoid was ramped linearly to full strength in 50 ms and run at full power for 1 s, 
them ramped down in a reverse of the ramp up procedure.
The beam polarization was then measured and the data in Fig.~3 in \cite{PolLifetime_PRE2003} were obtained,
for various settings of the rf solenoid frequency.
From the information in \cite{PolLifetime_PRE2003},
the beam revolution frequency was $f_c = 1\,677\,551$ Hz.
Because of the presence of the partial Type-3 Snake in the IUCF Cooler,
the resonant spin tune was shifted slightly from $1+G\gamma$ and 
the central resonant frequency was $f_{\rm res} = 1\,403\,002 \pm 14$ Hz.
The rms integrated field of the rf solenoid was stated to be $\int B\,dl = 0.7$ T-mm.
Given the beam kinetic energy is 270 MeV, I calculated the spin resonance strength induced by the rf solenoid to be
\bq
\label{eq:my_eps}
\varepsilon = \frac{1+G}{4\pi}\,\frac{\sqrt2 \int B\,dl}{pc/e} \simeq 1.9\times 10^{-5} \,.
\eq
I verified the correctness of this value by running Froissart-Stora simulations,
with the ring (and rf solenoid) parameters given in \cite{PolLifetime_PRE2003}.
I swept the rf solenoid frequency across the (fixed) spin precession frequency.
I obtained perfect agreement with the Froissart-Stora formula \cite{FS} for both the vector and tensor polarizations.
Hence the calculated value of $\varepsilon = 1.9\times 10^{-5}$ 
in eq.~\eqref{eq:my_eps}
is in agreement with the parameter values listed in the paper.
From the above value of $\varepsilon$, the HWHM resonance width is
\bq
\Gamma_{\rm HWHM} = \varepsilon\,f_c = 1.9\times10^{-5}\times 1677551 \simeq 32 \ \textrm{Hz}\,.
\eq
Fig.~3 in \cite{PolLifetime_PRE2003} displays a graph of the asymptotic polarization around the resonance center (Lorenztian curve).
From the figure, the observed HWHM resonance width is about 40 Hz, in agreement with my calculation above.
In fact, the same data were also displayed in a different paper (Fig.~2 in \cite{IUCF_first_spinflip_deuteron}).
This latter paper employed the same rf solenoid and reported results for the first spin flipping of a stored polarized deuteron beam.
It was stated in \cite{IUCF_first_spinflip_deuteron} that the FWHM resonance width was about 75 Hz, 
hence a HWHM width of about $37.5$ Hz, in agreement with the above findings.
In fact, the studies in \cite{IUCF_first_spinflip_deuteron}
reported excellent agreement with the Froissart-Stora formula \cite{FS}, 
with the following fitted values of the resonance strengths,
for the vector and tensor polarization
\begin{subequations}
\label{eq:res_str_data}
\begin{align}
\varepsilon_v &= (17.3 \pm 0.6) \times 10^{-6} \,,
\\
\varepsilon_t &= (17.7 \pm 0.5) \times 10^{-6} \qquad (P_{zz} = +1) \,,
\\
\varepsilon_t &= (16.3 \pm 0.3) \times 10^{-6} \qquad (P_{zz} = -2) \,.
\end{align}
\end{subequations}
For the tensor polarization, the two tensor polarization states are listed.
We see that all the values are consistent with each other, and with my theoretical calculation of $19 \times 10^{-6}$ in eq.~\eqref{eq:my_eps}.

In the second part of the studies reported in \cite{PolLifetime_PRE2003},
the rf solenoid was run continuously, 
and the vector and tensor polarizations were measured as functions of time.
This was done at three frequency settings, corresponding to frequency ofsets of
$\Delta f_{\rm sol} = 0.38$ kHz, $0.20$ kHz and $0.10$ kHz from the center of the spin resonance.
Fig.~4 in \cite{PolLifetime_PRE2003} displays a semi-log plot 
for $P_z$ and $P_{zz}$ as functions of the time $t$
for a rf solenoid frequency offset of $\Delta f_{\rm sol} = 0.38$ kHz.
The data were fitted using exponentials
$P_z(t) = P_z(0) e^{-t/\tau_v}$ and $P_{zz}(t) = P_{zz}(0) e^{-t/\tau_t}$
and the vector and tensor polarization decay lifetimes $\tau_v$ and $\tau_t$ were obtained.
Similar fits were made for the rf solenoid frequency offsets $\Delta f_{\rm sol} = 0.20$ kHz and $\Delta f_{\rm sol} = 0.10$ kHz.
The data for the vector and tensor polarization decay lifetimes are shown in Fig.~5 in \cite{PolLifetime_PRE2003}
(see also Fig.~6).
It was found that the decay lifetimes were in the ratio
\bq
\frac{\tau_v}{\tau_p} \simeq 1.9 \pm 0.2 \,.
\eq
However, standard angular momentum theory yields a ratio of $3$.

Reading through the paper \cite{PolLifetime_PRE2003},
it puzzled me that in Fig.~4 in \cite{PolLifetime_PRE2003},
i.e.~a rf solenoid frequency offset of $\Delta f_{\rm sol} = 0.38$ kHz,
the data indicated that both the vector and tensor polarizations decayed exponentially, to asymptotic values of zero.
The measurements spanned a time interval of 300 s.
The vector polarization decayed from about $0.48$ to about $0.3$.
The tensor polarization decayed from about $0.6$ to about $0.2$ (for $P_{zz}=+1$)
and from about $1.05$ to about $0.5$ (for $P_{zz}=-2$).
I say ``puzzled'' because $0.38$ kHz = $380$ Hz, 
hence the data in \cite{PolLifetime_PRE2003}
were indicating that an rf solenoid, operated continuously 
at a fixed frequency about $10\times$ the HWHM resonance width from the resonance center,
was depolarizing the beam to an asymptotic value of zero.

I contacted the authors of the paper \cite{PolLifetime_PRE2003} about the matter.
I received a very kind reply, but the reply did state in essence that an rf solenoid, operated continuously 
at a frequency offset of $10\times$ HWHM from the resonance center, causes a stored polarized deuteron beam to depolarize to an asymptotic value of zero.
I have no explanation as to how this can be.
An rf solenoid (or localized solenoid field) 
tilts the stable polarization away from the vertical direction
and the polarization settles down to a nonzero asymptotic value (away from the resonance center).
Many beam dynamics studies of stored polarized beams (at IUCF) have shown that this is so.

\bit
\item
I have fitted multiple examples of beam dynamics data in studies at the IUCF Cooler.
First, I was a coauthor on the paper reporting the results of the first spin dynamics studies at the UCF Cooler \cite{IUCF_first_test_1989}.
In this study, the solenoid fields in the cooling section were varied and the polarization was plotted as a function of the integrated field.
A localized static solenoid field is equivalent to an rf solenoid operated at an integer multiple of the beam revolution frequency, 
and the solenoid field was clearly operated continuously.
The initial (injected) polarization direction is vertical, but the stable polarization is not.
The polarization settles to a nonzero asymptotic value (away from the resonance center), pointing along a non-vertical direction.
Both the vertical and radial polarization components are plotted in Fig.~2 in \cite{IUCF_first_test_1989} 
and are fitted by the theory equations in the paper.
Those are my theory equations in the paper (eqs.~1-4), for the vertical and radial polarization components.
Actually, my theoretical analysis was sufficiently accurate that it revealed a sign error for the integrated solenoid field
in an early draft of the manuscript.

\item
Next, although this is not rf solenoids, a ``partial Type 3 Siberian Snake'' was found in the IUCF Cooler
\cite{Pollock_Type3}.
Pollock described the origin of the spin tune shift
due to noncommuting spin rotations in the electron cooling section, 
and gave a theoretical derivation of the spin tune shift.
(Notice that the theory curves in Fig.~2 in \cite{IUCF_first_test_1989} are slightly displaced from the data points.)
I also analyzed the theory equations of the partial Type 3 Snake for myself.
I found that the spin rotation angle was proportional to $G\gamma+1$, not $G\gamma$ as Pollock had written.
Since $G\gamma = 2$ in the studies reported in \cite{Pollock_Type3},
this changed the answer (spin tune shift) by a factor of $3/2$.
The spin rotation angle was 39 mrad, not 26 mrad.
(I forwarded a copy of my calculations to Pollock.)
The factor of $G\gamma+1$ which I calculated is now accepted as correct.
The paper by Minty and Lee \cite{Minty_Lee_Type3_1993},
which describes more extensive investigations of the partial Type 3 Snake,
states that the spin rotation angle was 39 mrad at a kinetic energy of 104.6 MeV
(the data which Pollock \cite{Pollock_Type3} and I analyzed).
I also realized that the excess spin rotation 
caused by the partial Type 3 Snake in the IUCF Cooler was an example of 
Berry's phase (actually an Aharanov-Anandan phase) or a topological quantum phase \cite{Mane_AAphase}.

\item
To return to rf solenoids,
in 1992 I saw a preprint from the University of Michigan,
describing spin resonances induced by an rf solenoid in the IUCF Cooler.
The spin resonances were induced by operating an rf solenoid (continuously) in the IUCF Cooler.
The stable polarization was again tilted away from the vertical direction.
A central resonance (well described by a Lorentzian) was observed, plus two narrower sideband resonances due to synchrotron oscillations.
I fitted that data in a BNL preprint (unpublished)
\cite{Mane_AGS364}.
The data was published in a 1998 paper:
see Fig.~5 in \cite{Krisch_1998}.
(The main part of that paper is actually about something else.)
However, the data had in fact been published earlier, 
in a 1996 paper by Lee and Berglund:
see Fig.~5 in \cite{Lee_Berglund_1996}.
(Krisch gave permission to Lee and Berglund to publish the data, 
acknowledged in \cite{Lee_Berglund_1996}.)
The formula by Lee and Berglund is very similar to mine from 1992 \cite{Mane_AGS364};
see the equations in the text below Fig.~5 in \cite{Lee_Berglund_1996}.
In terms of their notation, I obtained $g^2 \simeq 0.1$.  
Lee and Berglund obtained $g \simeq 0.3$ so $g^2 \simeq 0.09$, so they and I obtained almost the same fit to the data.
I fitted only the vertical polarization data (Fig.~5 in \cite{Krisch_1998}).
Lee and Berglund published and fitted the data for both the radial and vertical polarization components.
Actually the Lee-Berglund paper is the only place where the radial polarization data was published.
In any case, the data in \cite{Krisch_1998} and \cite{Lee_Berglund_1996}
demonstrated that an rf solenoid, operating continuously, at the IUCF Cooler, 
caused the polarization to settle to a nonzero asymptotic value (away from the resonance center) pointing in a non-vertical direction.
The width of the resonance (Lorentzian curve in the vertical polarization data) is given by the formula for the resonance strength for a localized rf solenoid.
The theory formulas for the central resonance are essentially those I mentioned earlier for the 1989 paper
\cite{IUCF_first_test_1989}.
There was no frequency sweep of the rf solenoid for the data in
Fig.~5 in \cite{Krisch_1998} and Fig.~5 in \cite{Lee_Berglund_1996};
the rf solenoid was operated continuously at a fixed frequency.
The rf solenoid frequency was set to various values and the asymptotic polarization was measured.
As I stated above, a central parent resonance and two synchrotron sideband resonances were observed.

\eit

\vfill\pagebreak
The data in Fig.~3 of the 2003 paper on the deuteron polarization lifetime data 
\cite{PolLifetime_PRE2003}
(Lorenztian curve, rf solenoid operated for 1 s) tell me that the spin components orthogonal 
to the stable polarization direction decohered and the deuteron polarization settled down to its asymptotic value on a timescale of less than 1 s.
Operating an rf solenoid continuously causes the polarization to settle down to a nonzero asymptotic value (away from the resonance center).
This is what was observed in studies with polarized protons at the IUCF Cooler (published in the Lee-Berglund paper \cite{Lee_Berglund_1996}).
I can explain such behavior, including the synchrotron sidebands.
Lee and Berglund fitted the same data independently, and obtained almost the same parameter fits as myself.
I fitted only the vertical polarization data, but Lee and Berglund fitted both the vertical and radial components.
My theoretical analysis of the data in the 1989 paper 
\cite{IUCF_first_spinflip_deuteron}
was sufficiently accurate that it revealed a sign error in 
the experimental parameters in an early draft of the manuscript.
I also derived the correct theoretical formula for the spin tune shift of the partial Type 3 Snake in the IUCF Cooler  
and I recognized the spin tune shift was a signature of a topological quantum phase
\cite{Mane_AAphase}.

The rf solenoid used in the deuteron polarization lifetime studies \cite{PolLifetime_PRE2003}
was also used in the spin-flipping studies of stored polarized deuteron beams in the IUCF Cooler
\cite{IUCF_first_spinflip_deuteron}
and yielded results in agreement with the theoretical calculations 
using  the Froissart-Stora formula \cite{FS}.
The experimentally fitted values for the resonance strengths
(see eq.~\eqref{eq:res_str_data})
and my calculated value
(see eq.~\eqref{eq:my_eps})
are in agreement.
These are all studies performed at the IUCF Cooler, with localized solenoid fields.

So, to answer Waldo MacKay: 
Ultimately, I have no explanation how an rf solenoid, operated continuously at a frequency offset $10\times$ HWHM from the center of a spin resonance, can depolarize a stored polarized beam to an asymptotic level of zero. 
And without such an explanation, I cannot explain the ratio of the polarization decay lifetimes of the tensor and vector polarizations for a beam of stored polarized deuterons.

\vskip 0.1in
{\em I could have said all this to Waldo MacKay ten years ago. It is my fault that I did not.}


\vfill\pagebreak

\end{document}